\documentclass{aa}

\usepackage{amsmath}
\usepackage{graphicx}
\usepackage{wasysym}

\begin{document}

\title{Properties of interstellar wind leading to shape morphology
of the dust surrounding HD 61005}

\author{P. P\'{a}stor}
\institute{Tekov Observatory, Sokolovsk\'{a} 21, 934~01 Levice,
Slovak Republic\\
\email{pavol.pastor@hvezdarenlevice.sk}}

\date{}

\abstract
{}
{A structure formed by dust particles ejected from the debris ring around
HD 61005 is observed in the scattered light. The main aim here is to constrain
interstellar wind parameters that lead to shape morphology in the vicinity
of HD 61005 using currently available observational data for the debris ring.}
{Equation of motion of 2 $\times$ 10$^5$ dust particles ejected from
the debris ring under the action of the electromagnetic radiation, stellar
wind, and interstellar wind is solved. A two-dimensional (2D) grid is placed
in a given direction for accumulation of the light scattered on the dust
particles in order to determine the shape morphology. The interaction
of the interstellar wind and the stellar wind is considered.}
{Groups of unknown properties of the interstellar wind that create
the observed morphology are determined. A relation between number densities
of gas components in the interstellar wind and its relative velocity
is found. Variations of the shape morphology caused by the interaction
with the interstellar clouds of various temperatures are studied.
When the interstellar wind velocity is tilted from debris ring axis
a simple relation between the properties of the interstellar wind and
an angle between the line of sight and the interstellar wind velocity
exists. Dust particles that are most significantly influenced by stellar
radiation move on the boundary of observed structure.}
{Observed structure at HD 61005 can be explained as a result of dust
particles moving under the action of the interstellar wind. Required
number densities or velocities of the interstellar wind are much higher
than that of the interstellar wind entering the Solar system.}


\titlerunning{Properties of ISW creating observed morphology at HD 61005}
\maketitle

\section{Introduction}
\label{sec:intro}

An amount of dust sufficient to produce significant infrared excess
was discovered at HD 61005 during the {\it Spitzer} FEPS survey.
The star was selected by \citet{hines} as a promising target for
{\it HST} NICMOS observations. In the scattered light, they resolved
an asymmetric structure of dust at the star. \citet{hines} suggested
that the observed morphology could be formed as a result of
an interaction between dust particles originating in the system and
a gas from an interstellar medium (ISM) moving with respect to the star
(interstellar wind). The fact that the star's proper motion roughly points
north \citep{leeuwen}, while the disk's swept-back ``wings'' are directed
to the south, might be more than just a coincidence \citep{hines,esposito}.
A result of the interaction was obtained numerically in \citet{debes}.
They used a debris ring ejecting dust particles on unbound orbits
varying further due to the interstellar wind (ISW) and the stellar
gravity reduced by Keplerian term from Poynting--Robertson (PR) effect.
\citet{maness} used dust particles on bound orbits interacting with
the ISM gas in order to obtain the observed morphology. Their
results were based on a secular variation of dust orbits
due to the ISM gas taken from \citet{scherer}. According
to results in \citet{scherer} the dust particles on bound orbits
should undergo an increase of the semimajor axis due to the acceleration
from the ISW. The increase of the semimajor axis
should create the observed swept-back structure in \citet{maness}.
However, \citet{flow} and \citet{dyncd} analytically proved
that for the dust particles on bound orbits, the semimajor axis always
decreases independently of an orientation of the orbit with respect
to the interstellar gas velocity vector.

The debris ring ejecting the dust particles was resolved in the scattered
light by \citet{buenzli}. According to their observations, the bulk
of the debris ring with low eccentricity is roughly located between 55 and
65 AU from HD 61005 with an apparent inclination of 5.7 from edge-on.
\citet{esposito} suggested that dust asymmetry observed at HD 61005
can be explained as a result of secular perturbations of dust
particles on bound orbits by an eccentric and inclined planet. Such
an explanation requires smooth boundaries at the ends of the swept-back
structure (due to the bound orbits), which are not observed as far
as 350 AU from the star \citep{schneider}. Moreover, scattering behavior
is more consistent with smaller particles below blow-out radius \citep{hines}.

In this study we use observed properties of the debris ring at HD 61005
in order to explain observed asymmetric morphology using interaction
of the dust particles on unbound orbits with the ISW.
We concentrate on groups of unknown properties of the ISM gas
that lead to the observed morphology. We develop \citet{debes} results,
using the observed dimensions of the debris ring and using
correct acceleration of the dust particles caused by the ISW.

\section{Model}
\label{sec:model}

HD 61005 is the G8V star with visual magnitude 8.22 at 35.4 pc.
Its position and kinematics with respect to the Sun are typical for
members of the Argus association \citep{desidera}. Stellar age
of the Argus association is believed to be 40 Myr \citep{torres}.
The optical and near infrared (IR) photometry measurements suggest that the
star has a temperature similar to our Sun. In far-IR the star shows IR-excess
from the debris disk observed also in the scattered light
\citep[e.g.,][]{schneider}. We verified stellar parameters from
\citet{olofsson} and use them in this study. The luminosity and radius
of the star used are $L_{\star}$ $=$ 0.58 $L_{\odot}$ and
$R_{\star}$ $=$ 0.84 $R_{\odot}$, respectively.
For the age 40 Myr and an effective temperature of 5500 K
(used to scale the photometry with Kurucz library) isochrones
from \citet{siess} give the stellar mass 1.1 $M_{\odot}$.
Thus, HD 61005 is considered as a pre-main sequence star.

\subsection{Electromagnetic radiation}
\label{subsec:elemag}

The acceleration of moving spherical dust particles caused by
the electromagnetic radiation is the PR effect
\citep{poynting,robertson,PRI,CMDA,icarus}. The acceleration is
\begin{equation}\label{PR}
\frac{d \vec{v}}{dt} = \frac{\mu \beta}{r^{2}}
\left [ \left ( 1 - \frac{\vec{v} \cdot \vec{e}_{\text{R}}}{c} \right )
\vec{e}_{\text{R}} - \frac{\vec{v}}{c} \right ] ~,
\end{equation}
where $\mu$ $=$ $G M_{\star}$, $G$ is the gravitational constant,
$M_{\star}$ is the mass of the star, $r$ is the stellocentric distance,
$\vec{e}_{\text{R}}$ is the radial unit vector directed from the star
to the dust particle, $\vec{v}$ is the velocity of the particle with
respect to the star, and $c$ is the speed of light. The parameter $\beta$
is defined as the ratio between the electromagnetic radiation pressure
force and the gravitational force between the star and the particle
at rest with respect to the star
\begin{equation}\label{beta}
\beta = \frac{L_{\star} \bar{Q}'_{\text{pr}} A'}{4 \pi c \mu m} ~.
\end{equation}
Here, $L_{\star}$ is the stellar luminosity, $\bar{Q}'_{\text{pr}}$ is
the dimensionless efficiency factor for the radiation pressure averaged
over the stellar spectrum and calculated for the radial direction
($\bar{Q}'_{\text{pr}}$ $=$ 1 for a perfectly absorbing sphere),
$c$ is the speed of light in vacuum, and $A'$ is the geometric
cross-section of the particle with mass $m$.

\subsection{Mass-loss rate}
\label{subsec:MLR}

The solar heliosphere is created in an interaction of the solar wind
with the ISM gas incoming to the Solar system. Also in
the vicinity of HD 61005 its stellar wind will be influenced by
an incoming ISM gas. Far outside of the astrosphere
of HD 61005 we may neglect an acceleration of dust particles
caused by the stellar wind \citep[e.g.,][]{richardson}. For a correct
description of the stellar wind at HD 61005 we need to also know
the mass loss rate of this star. Recent results in \citet{johnstone}
enable us to approximately determine mass loss rate for a star of mass
between 0.4 $M_{\odot}$ and 1.1 $M_{\odot}$ using the angular
velocity of surface rotation, the radius, and the mass of the star.
Unfortunately, in their sample, the authors only considered stars
older than 100 Myr to fit parameters, although their theory also enables
consideration of the pre-main sequence stars. The relation
for the wind torque in \citet{matt} was originally derived also
for pre-main sequence stars and did not lose this applicability.
Also the relation for dependence of the X-ray flux as a function
of the stellar mass and the rotation period from \citet{wright}
is applicable for younger stars. Improvement of their fit for younger
stars is beyond the scope of this study. In order to determine the mass-loss
rate for HD 61005 we will simply assume that the fit obtained in
\citet{johnstone} can be used for this star. We will see later that
the astrosphere of HD 61005 is far inside the debris ring also
for much larger mass-loss rates and stellar-wind speeds.
Stellar rotation with period $P_{\star}$ $=$ 5.04 days has been
determined for HD 61005 in \cite{desidera} from photometric measurements.
This corresponds to the angular velocity of surface rotation
$\Omega_{\star}$ = $2 \pi / P_{\star}$ $\approx$ 1.4 $\times$ 10$^{-5}$
rad/s. For stable stars rotating with higher angular velocity than
a saturation angular velocity the mass-loss rate cannot be higher
\citep[e.g.,][]{pallavicini,wright}. The saturation angular velocity
for HD 61005 can be determined from Eq. 6 in \citet{johnstone} as
$\Omega_{\text{sat}}$ $\approx$ 5.0 $\times$ 10$^{-5}$ rad/s.
Since the obtained angular velocity of rotation for HD 61005
is below the saturation limit, the star is in an unsaturated
regime. The angular velocity of surface rotation
coupled with $R_{\star}$ $=$ 0.84 $R_{\odot}$ and
$M_{\star}$ $=$ 1.1 $M_{\odot}$ gives, using Eq. 4 in \cite{johnstone},
the mass loss rate $\dot{M}_{\star}$ $\approx$ 4.8 $\dot{M}_{\odot}$.

\subsection{Stellar wind}
\label{subsec:stellarwind}

Impinging particles of the stellar wind cause an acceleration
of the dust particle that can be determined from a covariant equation
of motion \citep{covsw}
\begin{equation}\label{SW}
\frac{d \vec{v}}{dt} = \frac{\mu \beta \eta u}{c \bar{Q}'_{\text{pr}} r^{2}}
\left [ \left ( 1 -
\frac{\vec{v} \cdot \vec{e}_{\text{R}}}{u} \right )
\vec{e}_{\text{R}} - \frac{\vec{v}}{u} \right ] ~,
\end{equation}
where $u$ is the speed of the stellar wind, $\eta$ is the ratio of the stellar
wind energy to the stellar electromagnetic radiation energy, both radiated
per unit time. For $\eta$ we can write
\begin{equation}\label{eta}
\eta = \frac{\dot{M}_{\star} c^{2}}{L_{\star}} ~.
\end{equation}
Since HD 61005 is in the unsaturated regime for the speed of the stellar
wind, we obtain, according to Fig. 13 in \citet{johnstone}, $u$ $\approx$
1200 km/s. Using the mass loss rate of HD 61005 we obtain $\eta$ $\approx$
2.4. Now, if we compare the Keplerian terms in Eqs. (\ref{PR}) and (\ref{SW}),
we can neglect the Keplerian term in Eq. (\ref{SW}) with
respect to the Keplerian term in Eq. (\ref{PR}) for dust particles with
$\bar{Q}'_{\text{pr}}$ $=$ 1.

\subsection{ISW}
\label{subsec:ISW}

Asymmetric structure observed at HD 61005 is commonly explained as result
of an ISM gas affecting the dynamics of dust particles originating
in the system \citep[e.g.,][]{debes,maness}. The acceleration used
in \citet{debes} was adopted from meteor theory and therefore has
different handling with drag coefficients not very suitable for the ISM gas.
In this work we use the correct form of acceleration causing the ISM gas
to influence the dynamics of a spherical particle. The acceleration
of the spherical dust moving through gas is \citep{baines}
\begin{equation}\label{ISM}
\frac{d \vec{v}}{dt} = - \sum_{i = 1}^{N} c_{\text{D}i} \gamma_{i}
\vert \vec{v} - \vec{v}_{\text{F}} \vert
\left ( \vec{v} - \vec{v}_{\text{F}} \right ) ~.
\end{equation}
The sum in Eq. (\ref{ISM}) runs over all particle species $i$.
$\vec{v}_{\text{F}}$ is the velocity of the ISM with
respect to the star. $c_{\text{D}i}$ in Eq. (\ref{ISM}) is the drag
coefficient
\begin{align}\label{cd}
c_{\text{D}i}(s_{i}) = {} & \frac{1}{\sqrt{\pi}}
      \left ( \frac{1}{s_{i}} + \frac{1}{2 s_{i}^{3}} \right )
      \text{e}^{-s_{i}^{2}} +
      \left ( 1 + \frac{1}{s_{i}^{2}} - \frac{1}{4 s_{i}^{4}} \right )
      \text{erf}(s_{i})
\notag \\
& + \left ( 1 - \delta_{i} \right )
      \left ( \frac{T_{\text{d}}}{T_{i}} \right )^{1 / 2}
      \frac{\sqrt{\pi}}{3s_{i}} ~,
\end{align}
where erf$(s_{i})$ is the error function $\text{erf}(s_{i})$ $=$
$2 / \sqrt{\pi} \int_{0}^{s_{i}} \text{e}^{-t^{2}} dt$, $\delta_{i}$ is
the fraction of impinging particles specularly reflected at the surface
(a diffuse reflection is assumed for the rest of the particles, see
\citealt{baines,gustafson}), $T_{\text{d}}$ is the temperature of the dust
grain, and $T_{i}$ is the temperature of the $i$th gas component.
$s_{i}$ in Eq. (\ref{cd}) is the molecular speed ratio
\begin{equation}\label{s}
s_{i} = \sqrt{\frac{m_{i}}{2 k T_{i}}} U ~.
\end{equation}
Here, $m_{i}$ is the mass of the atom in the $i$th gas component,
$k$ is Boltzmann's constant, and
$U$ $=$ $\vert \vec{v} - \vec{v}_{\text{F}} \vert$
is the relative speed of the dust particle with respect to the gas.
For the collision parameter $\gamma_{i}$ in Eq. (\ref{ISM}) we obtain
\begin{equation}\label{cp}
\gamma_{i} = n_{i} \frac{m_{i}}{m} A' ~,
\end{equation}
where $n_{i}$ is the number density of the $i$th kind of interstellar
atom. We see in Sect. \ref{subsec:interaction} that in the case
of HD 61005, the ISW velocity and number densities are not affected by
an interaction of the ISW with the stellar wind.

\subsection{Equation of motion}
\label{subsec:EOM}

The equation of motion, which determines the dynamics of dust particles in
an orbit around a star under the action of the electromagnetic radiation,
the stellar wind, and ISW are calculated based on Eqs. (\ref{PR}),
(\ref{SW}) and (\ref{ISM})
\begin{align}\label{EOM}
\frac{d \vec{v}}{dt} = {} & - \frac{\mu}{r^{2}}
      \left ( 1 - \beta \right ) \vec{e}_{\text{R}} -
      \frac{\mu \beta}{r^{2}}
      \left ( 1 + \frac{\eta}{\bar{Q}'_{\text{pr}}} \right )
      \left ( \frac{\vec{v} \cdot \vec{e}_{\text{R}}}{c}
      \vec{e}_{\text{R}} + \frac{\vec{v}}{c} \right )
\notag \\
& - \sum_{i = 1}^{N} c_{\text{D}i} \gamma_{i}
      \vert \vec{v} - \vec{v}_{\text{F}} \vert
      \left ( \vec{v} - \vec{v}_{\text{F}} \right ) ~.
\end{align}
Here also the approximation mentioned after Eq. (\ref{eta}) was considered.

\subsection{Distance to bow shock}
\label{subsec:BS}

The supersonic motion of a star producing a stellar wind in the local
ISM creates the so-called bow shock around a direction of the star velocity
in the ISM (upstream direction). The stellocentric distance to the bow shock
is minimal in the upstream direction. The minimal stellocentric distance
to the bow shock (stand-off distance) can be calculated from an equilibrium
between ram pressures of the stellar wind and the ISM \citep{wilkin}. We have
\begin{equation}\label{BS}
r_{\text{BS}~0} = \sqrt{\frac{\dot{M}_{\star} u_{\text{sw}}}
{4 \pi \varrho_{\text{F}} v_{\text{F}}^{2}}} ~,
\end{equation}
where $\varrho_{\text{F}}$ $=$ $\sum_{i = 0}^{N} n_{i} m_{i}$ is
the density of the ambient ISM. In an ideal case, the shape of the bow
shock is rotationally symmetric with an axis of the symmetry going
through the star in the direction of the ISW velocity. An analytical solution
for the bow shock shape that is consistent with numerical solutions
of magnetohydrodynamical (MHD) equations was derived in \citet{wilkin}
using the assumption of momentum conservation and assuming that the material
mixes and cools instantaneously (the thin shell approximation).
The derived solution determines the distance to the bow shock
$r_{\text{BS}}$ as a function of a polar angle $\theta$ measured
from a bow shock surface element with the unknown distance
to the upstream direction
\begin{equation}\label{rBS}
r_{\text{BS}} = r_{\text{BS}~0} \csc \theta
\sqrt{3 \left ( 1 - \theta \cot \theta \right )} ~,
\end{equation}
According to Eq. (\ref{rBS}) there is no difference in the shape of the bow
shocks and their size is determined only by the stand-off distance.

\subsection{Initial conditions}
\label{subsec:initial}

We assume that the source of dust particles forming the swept-back
structure at HD 61005 is in the circumstellar ring observed also in
the scattered light. As the source of dust particles, we consider parent
bodies in elliptical orbits with semimajor axes distributed
randomly between 55 and 65 AU. We assume that the elliptical orbits
have randomly distributed eccentricities smaller than 0.1.
Inclinations of the orbits are distributed randomly between
0 and 2$^{\circ}$. Arguments of pericenters, longitudes of ascending
nodes, and true anomalies of the parent bodies are distributed randomly
between 0 and 2$\pi$. The dust particle, after ejection from
the parent body, is subjected to the radiation of the star. The position
and velocity in a moment of the ejection can be used in order to calculate
initial orbital elements of the dust particle. After the ejection, central
Keplerian acceleration from the star is reduced by a factor $1 - \beta$
(see text after Eq. \ref{eta}) and the particle obtains a different
oscular orbit. The observed swept-back structure at HD 61005 can be explained
with dust particles moving in hyperbolic orbits. The dust particles in
the hyperbolic orbits have also been observed in the Solar system and are
called $\beta$-meteoroids \citep{meteoroids}. Calculation of initial
orbital elements after an ejection to the hyperbolic orbit is shown in
Appendix \ref{appendix}.

\begin{figure*}
\begin{center}
\includegraphics[width=0.8\textwidth]{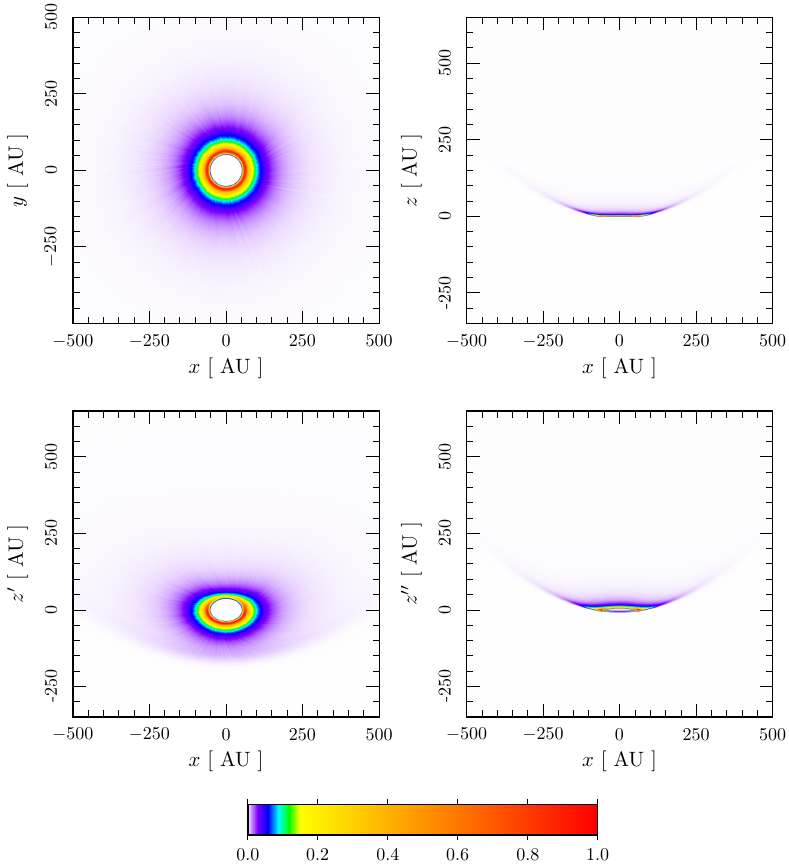}
\end{center}
\caption{An interaction of dust particles with $R$ $=$ 0.168, $\varrho$ $=$
1 g.cm$^{-3}$ and $\bar{Q}'_{\text{pr}}$ $=$ 1.663 ejected from parent bodies
in the debris ring around HD 61005 located between 55 and 65 AU and
the interstellar wind with number densities $n_{\text{H}}$ $=$ 80 cm$^{-3}$,
$n_{\text{He}}$ $=$ 6 cm$^{-3}$, and $n_{\text{p}}$ $=$ 12.8 cm$^{-3}$
moving perpendicularly to the debris ring with relative velocity 30 km/s
with respect to the star. Color-scale is linear with observed intensity
of light with wavelength $\lambda$ $=$ 580 nm scattered on
the dust particles according to Mie scattering theory. The ring mid-plane
is located in the $xy$-plane. The bottom-left plot is created by a rotation
of a viewpoint in the top-left plot by the angle 45$^{\circ}$ counterclockwise
around $x$-axis in $yz$-plane. The bottom-right plot is created by
the rotation of a viewpoint in the top-left plot by the angle 95.7$^{\circ}$
counterclockwise around the $x$-axis in $yz$-plane in order to show
the shape morphology observed from the Earth.}
\label{fig:usable}
\end{figure*}

Properties of the star HD 61005 are sufficiently well known. Unfortunately,
this cannot be said about properties of the ISM gas which varies the dynamic
of the observed dust. After a few numerical solutions of the equation of motion,
it is easy to see that observed structure requires the acceleration caused
by the ISM to be dominant in the dynamics of ejected dust particles.
The dominance can be accomplished by high density, high speed and/or
high temperature of the ISM gas. The observed shape morphology
is obtained for the ISW with various velocities and densities
of gas components. The high temperature possibility alone
requires ISM gas temperatures of more than 10$^{5}$ K and for such high
temperatures at HD 61005 there is no observational evidence. Therefore,
the gas temperature will not be considered as the primary cause for the
observed morphology. The influence of the ISW with various temperatures
on the shape morphology will discussed separately. Implications for
the shape morphology from the specular as well as the diffuse reflection
at the surface of the dust grains will be compared.

\subsection{Observations in scattered light}
\label{subsec:observations}

We numerically solved the equation of motion for 2 $\times$ 10$^{5}$
particles. Obtained trajectories were accumulated into a 2D grid observed
from a given direction. A similar technique was used also by \citet{debes}
and earlier for visualization of resonant structures in debris
disks by \citet{LZ1999}, \citet{MM}, \citet{SK} and others.
This technique enables visualization of an optical depth in a given
element of the grid. In order to compare our model with observations of dust
in the scattered light, we calculated an intensity of light from HD 61005
scattered on each dust particle in the given element of the grid in
the observed direction using Mie scattering theory \citep[e.g.,][]{mie,hulst}
with an assumption that the light travels through optically thin environment.
Mie theory represents one solution of Maxwell's equations for
boundary conditions used in our model.

\begin{figure}[t]
\begin{center}
\includegraphics[width=0.44444444444444444444444444444444\textwidth]{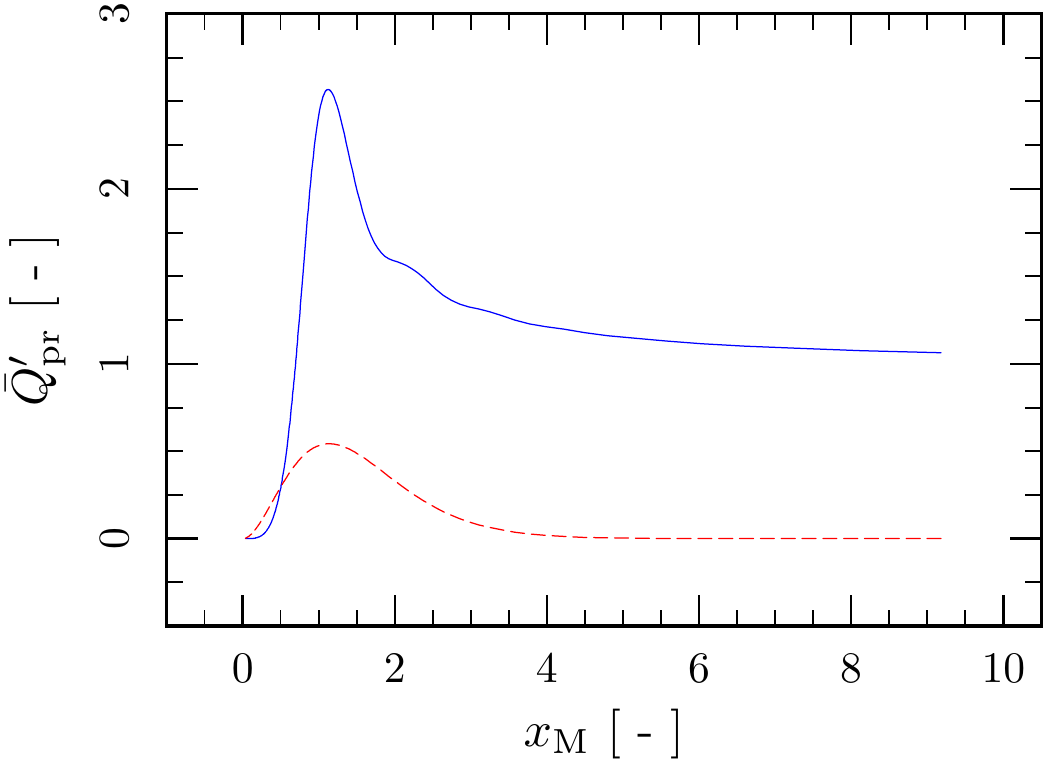}
\end{center}
\caption{Dimensionless efficiency factor for the radiation pressure
($Q'_{\text{pr}}$) determined for totally reflecting particles (solid line).
Spectral distribution of Planck's blackbody with temperature 5500 K in
dependence on $x_{\text{M}}$ $=$ $2 \pi R / \lambda$ for $R$ $=$ 0.168 $\mu$m
is shown for comparison (dashed line). The maximal $\bar{Q}'_{\text{pr}}$
is obtained for the dust particle with $R$ $=$ 0.168 $\mu$m.}
\label{fig:Qpr}
\end{figure}

\section{Results}
\label{sec:results}

By numerically solving the equation of motion one can easily find
various properties of the ISM that lead to practically identical
shape morphology. It is impossible using existing observational
data to decide which properties really belong to the ISM gas at HD 61005.
Instead of choosing only one property, we focus on groups
of properties that give the observed structure using our dynamical
model. The relative speed of HD 61005 with respect to the Sun determined from
its measured proper motion is 27.0 km/s. The ISM in the Solar System
arrives from the direction $\lambda_{\text{ecl}}$ $=$ 254.7$^{\circ}$
(heliocentric ecliptic longitude) and $\beta_{\text{ecl}}$ $=$
5.2$^{\circ}$ (heliocentric ecliptic latitude) and moves with
a speed $v_{\text{F}}$ $=$ 26.3 km/s \citep{lallement}. If the Sun
and the HD 61005 were embedded in the same interstellar cloud,
the speed of the ISM gas at HD 61005 would be only 12.8 km/s. The ISM
with such a small speed requires large densities of the ISM gas in
order to create the observed morphology (see further). Therefore, it is
more probable that the Sun and the HD 61005 are not embedded in the same
interstellar cloud.

\begin{figure}[t]
\begin{center}
\includegraphics[width=0.28801169590643274853801169590643\textwidth]{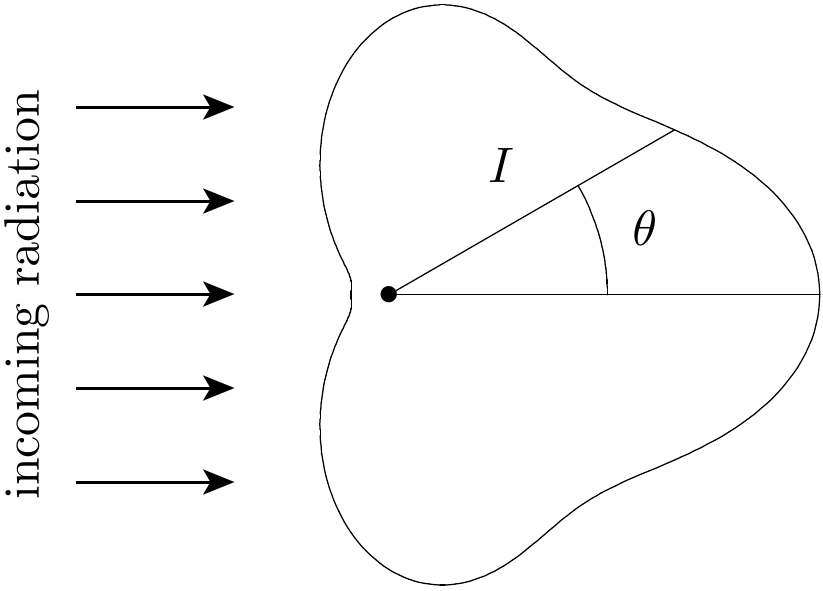}
\end{center}
\caption{Scattering diagram of light with $\lambda$ $=$ 580 nm impinging on
the dust particle with $R$ $=$ 0.168 $\mu$m. This particle gives the maximal
dimensionless efficiency factor for the radiation pressure averaged
over the stellar spectrum. For $x_{\text{M}}$ $=$ $2 \pi R / \lambda$
$<$ 1.38 scattering is predominantly back to the source. Considered
situation corresponds to $x_{\text{M}}$ $\approx$ 1.820 and the scattering
diagram has the usual preponderance of forward radiation over back
radiation \citep{hulst}.}
\label{fig:diagram}
\end{figure}

\subsection{ISW velocity perpendicular to the debris ring plane}
\label{subsec:perpendicular}

In Fig. \ref{fig:usable} is depicted the obtained shape
morphology consistent with the observed morphology at HD 61005.
The interstellar gas velocity is perpendicular to the debris ring.
In order to reproduce the observed morphology we used number
densities for hydrogen, helium, and plasma in the same ratios as number
densities of the ISM incoming to the Solar system. We have adopted
the following number densities $n_{\text{H~S}}$ $=$ 0.2 cm$^{-3}$,
$n_{\text{He~S}}$ $=$ 0.015 cm$^{-3}$, and $n_{\text{p~S}}$ $=$ 0.032 cm$^{-3}$
from \citet{frisch} for interstellar hydrogen, helium, and plasma entering
the Solar system, respectively. Temperatures of all ISM gas components
were equal to $T$ $=$ 6300 K, which is approximately the temperature
of interstellar helium entering the Solar system that is a gas component
weakly affected in the interaction of the ISW with the stellar wind.
We assumed that the ISM atoms are specularly reflected at the surfaces
of the dust grains ($\delta_{i}$ $=$ 1 in Eq. \ref{cd}).
Chosen speed of the ISM is $v_{\text{F}}$ $=$ 30 km/s and
the number densities found are $n_{\text{H}}$ $=$ 80 cm$^{-3}$,
$n_{\text{He}}$ $=$ 6 cm$^{-3}$, and $n_{\text{p}}$ $=$ 12.8 cm$^{-3}$
for interstellar hydrogen, helium, and plasma, respectively.
 Cartesian coordinates are used in the plots with the origin
in the star. The bottom-left plot is created by a rotation
of a viewpoint in the top-left plot by 45$^{\circ}$
counterclockwise around the $x$-axis in the $yz$-plane. The bottom-right
plot is created by the rotation of a viewpoint in the top-left plot by
95.7$^{\circ}$ counterclockwise around the $x$-axis in the $yz$-plane in order
to show the shape morphology observed from the Earth \citep{buenzli}.
The dominance of the acceleration caused by the ISM gas enables
us to use only single particle size for a creation of correct morphology shape
for the given ISM parameters. In this case the shape of swept-back structure
is not a function of the particle's radius $R$. This can also be proved
analytically using a simplified approximation that the acceleration
of the dust particle sufficiently far from the star is given only by
the Stark approximation. In the Stark approximation, the acceleration caused
by the ISM gas depends on the velocity as $C v_{\text{F}} \vec{v}_{\text{F}}$,
where $C$ is a constant \citep[see][]{fgf}. However, the shape
of swept-back structure is a function of $\bar{Q}'_{\text{pr}}$. The boundary,
determining the curvature of ``wings'', is formed by dust particles that
are most significantly influenced by the stellar radiation; hence,
the particles with maximal $\bar{Q}'_{\text{pr}}$ obtained from Mie
scattering theory. These particles obtain maximal variation of the velocity
in approximately radial direction close to the star. The maximal
$Q'_{\text{pr}}$ is obtained for totally reflecting particles.
For the effective temperature of the star 5500 K we can
calculate the particle's radius for which $\bar{Q}'_{\text{pr~max}}$
is obtained. The obtained radius is 0.168 $\mu$m and corresponding
maximal $\bar{Q}'_{\text{pr}}$ is 1.663 (Fig. \ref{fig:Qpr}). A slightly
larger radius of about 0.2 $\mu$m for the scattering particles has been
suggested from observations by \citet{hines}. For this radius we
obtain $\bar{Q}'_{\text{pr}}$ $=$ 1.642, not far from
the $\bar{Q}'_{\text{pr~max}}$. Fig. \ref{fig:usable} was obtained
using icy dust particles with radius $R$ $=$ 0.168 $\mu$m.
 In Eq. (\ref{beta}) these parameters give $\beta$ $\approx$ 3
for the stellar parameters of HD 61005. Scattering intensities
were obtained at $\lambda$ $=$ 580 nm. This wavelength
is free of any major Fraunhofer absorption lines in the stellar
spectrum. The positions of dust particles are not dependent
on wavelength of the observed scattered light. Therefore, the shape
morphology should not depend on the chosen wavelength of scattered
light in narrow visible light wavelength interval. Scattering intensities
obtained using Mie theory should be exactly correct for the used model.
Scattering diagram for the used particle at 580 nm is shown in
Fig. \ref{fig:diagram}.

\begin{figure}[t]
\begin{center}
\includegraphics[width=0.44005847953216374269005847953216\textwidth]{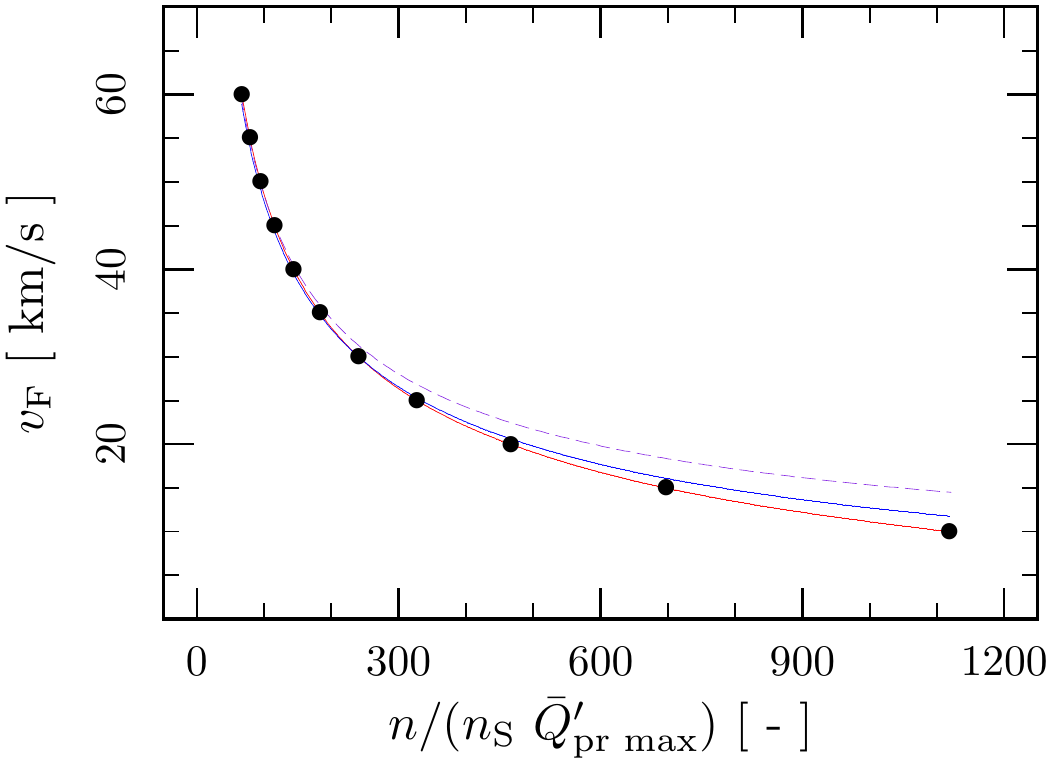}
\end{center}
\caption{Relation between the interstellar gas speed and the component
ratios of the interstellar gas number densities at HD 61005 and the Sun
(red lines) that give the observed morphology at HD 61005. Used interstellar
gas number densities at the Sun are $n_{\text{H~S}}$ $=$ 0.2 cm$^{-3}$,
$n_{\text{He~S}}$ $=$ 0.015 cm$^{-3}$,and $n_{\text{p~S}}$ $=$ 0.032 cm$^{-3}$.
After division of the density ratios by maximal averaged efficiency
factor for the radiation pressure, we obtain the relation independent
of the particle properties in our model. The blue line shows
the Stark approximation with considered dependence of the drag coefficients
on the velocity of the ISM gas. The dashed line shows a function
$v_{\text{F}}$ $=$
$C_{\text{f}} \sqrt{n_{\text{S}}~\bar{Q}'_{\text{pr~max}} / n}$,
where $C_{\text{f}}$ is fitted to the measured data.}
\label{fig:density}
\end{figure}

\subsection{Speeds and densities of the ISW}
\label{subsec:density}

The acceleration caused by the ISM gas has such a form that smaller
velocity and larger densities of the ISM gas can give the same resulting
shape morphology as a larger velocity and smaller densities.
In Fig. \ref{fig:density} are shown velocity and number densities of the ISM
that produce the observed shape of the swept-back structure at HD 61005
using presented model for the dust particles with a given
$\bar{Q}'_{\text{pr~max}}$. The number densities are shown as component
density ratios of the interstellar matter at HD 61005 and the interstellar
matter at the Sun. Temperatures of all ISM gas components were
equal to $T$ $=$ 6300 K. Different temperatures will be taken into account
in Sect. \ref{subsec:temperature}. We have used the specular reflection
at the surface of the dust grains ($\delta_{i}$ $=$ 1 in Eq. \ref{cd}).
Variations of the shape morphology caused by the diffuse reflection
at the surface can be neglected (see Sect. \ref{subsec:diffusion}).
The red line is obtained from numerical solution of the equation
of motion (Eq. \ref{EOM}) using a spline interpolation
applied on the measured values (black circles). The shape depicted in
Fig. \ref{fig:usable} corresponds to the component density ratio
$n / ( n_{\text{S}}~\bar{Q}'_{\text{pr~max}} )$ $\approx$ 240 on the red
line. The observed shape morphology can be obtained with the density ratios
smaller than 240 for dust particles with $\bar{Q}'_{\text{pr~max}}$ $<$ 1
at the ISW speed 30 km/s. The blue line shows the Stark approximation
$\sum_{i = 1}^{3} c_{0i} n_{i} m_{i}
v_{\text{F}}^{2} / \bar{Q}'_{\text{pr~max}}$ $=$
constant $\approx$ 1.1567 $\times$ 10$^{-10}$ kg.m$^{-1}$.s$^{-2}$, where
$c_{0i}$ $=$ $c_{Di}(s_{0i})$ and $s_{0i}$ $=$
$\sqrt{m_{i} / 2 k T_{i}} ~v_{\text{F}}$. The dependence of $c_{\text{D}}$
on the speed of the ISM gas ($v_{\text{F}}$) is considered in
this approximation. The constant was calculated from the case
depicted in Fig. \ref{fig:usable} ($v_{\text{F}}$ $=$ 30 km/s).
In the Stark approximation $c_{\text{D}}$ does not depend on the velocity
of the dust particle. If $c_{\text{D}}$ were constant for
all the ISW speeds, then the approximative relation would be
$v_{\text{F}}$ $=$
$C_{\text{f}} \sqrt{n_{\text{S}}~\bar{Q}'_{\text{pr~max}} / n}$,
where $C_{\text{f}}$ is a constant. Such a function is depicted in
Fig. \ref{fig:temperature} with the dashed line. The solution
of the equation of motion is approximated fairly well by both
simplified relations. The plots depicting the simplified
relations were obtained only with the acceleration from the ISM gas.
This demonstrates the above mentioned dominance of the ISM gas
in the dynamical evolution of the dust particles farther from the star.
The star's gravity and the PR effect can by neglected in comparison
with influence of the ISM gas farther from the star.

In \citet{debes}, a speed of 25 km/s was used for the interstellar gas. Using
depicted results in Fig. \ref{fig:density} we obtain the same observed
morphology for dust particles with $\bar{Q}'_{\text{pr~max}}$ $=$ 1 at
number densities $n_{\text{H}}$ $\approx$ 65.4 cm$^{-3}$, $n_{\text{He}}$
$\approx$ 4.9 cm$^{-3}$, and $n_{\text{p}}$ $\approx$ 10.5 cm$^{-3}$.
The model picture in \citet{debes} was obtained with $n_{\text{H}}$
100 cm$^{-3}$. We found that such high densities of hydrogen alone
are not necessary for three reasons: 1) \citet{debes} used
an assumption ``that the grains originate from radii $\sim$10 AU from
the star'' (see p. 324 therein). In 2009 shape of the debris ring
was not yet determined. The parent bodies on such orbits produce
dust particles with much larger initial velocities than the parent
bodies in the observed debris ring between 55 and 65 AU. The dust
particle moving with such a large initial tangential velocity requires
much intensive acceleration from the ISW to be pushed up
from the ring plane sufficiently fast. 2) The acceleration in \citet{debes}
has a different form of the drag coefficient. For the same dust particle,
the same hydrogen density, and the relative particle-gas speed 25 km/s, we
obtain that the acceleration in this paper is $\sim$1.40 times larger
than the acceleration in \citet{debes} at the temperature of hydrogen
6300 K. 3) It is very probable that gas components in addition to hydrogen
exist in the interstellar matter at HD 61005.

If the curvature of swept-back structure were observed farther from
the star, then it should be possible to determine densities of ISM components
directly. The data in Fig. \ref{fig:density} were fitted to match available
observations from \citet{schneider} to distance 230 AU along the ring mid-plane
with correct distance scaling from \citet{buenzli}.

\begin{figure}[t]
\begin{center}
\includegraphics[width=0.44005847953216374269005847953216\textwidth]{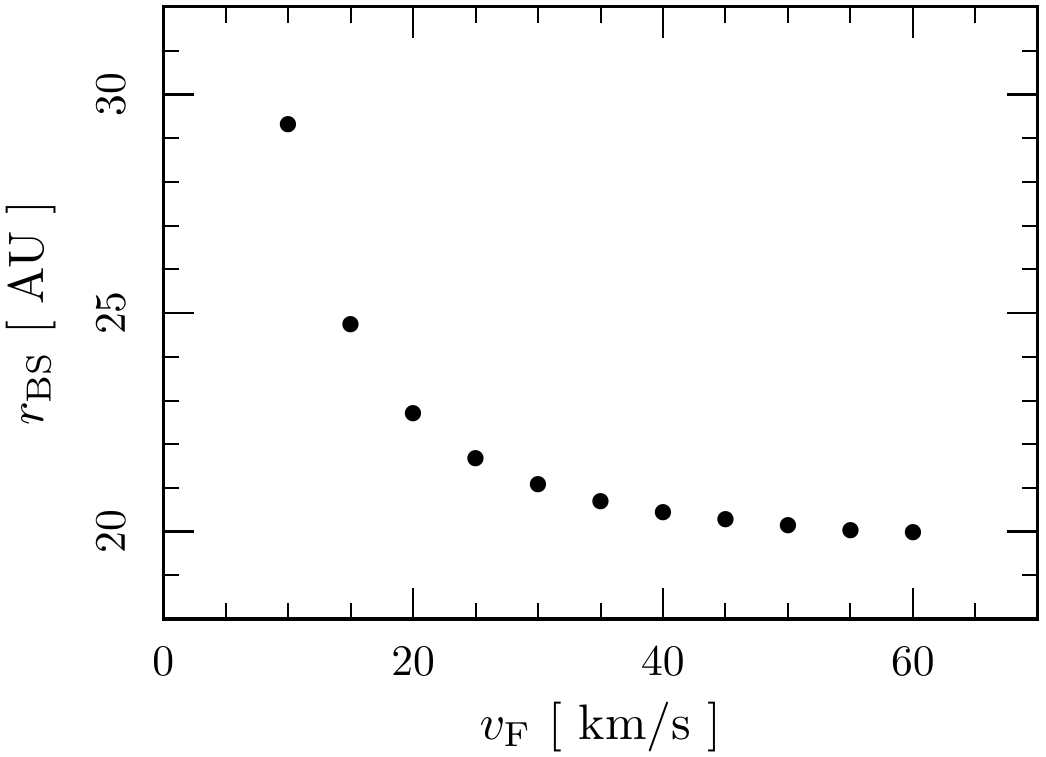}
\end{center}
\caption{Distances to the bow shock of the ISW and the stellar wind
interaction in the debris ring mid-plane. The distances were obtained
using Eq. (\ref{rBS}) with $\theta$ $=$ 90$^{\circ}$ for the ISW
parameters giving the observed shape morphology for the dust particles
with $\bar{Q}'_{\text{pr~max}}$ $=$ 1.663 (see Fig. \ref{fig:density}).
Calculated distances to the bow shock are significantly smaller than
the stellocentric distances 55-65 AU at which the bulk of the debris
ring ejecting the dust particles is located. Therefore, the debris
ring should not be located in the bow shock. Since, the termination
shock of the stellar wind is closer to the star than the bow shock,
the dynamics of dust particles should not be affected by the
stellar wind.}
\label{fig:BS}
\end{figure}

\subsection{Interaction of the ISW with stellar wind}
\label{subsec:interaction}

For the ISW gas considered in Fig. \ref{fig:density},
the interaction of the ISW with the stellar wind should be supersonic
\citep[see e.g.,][]{holzer}. Hence, the bow shock should be formed.
Even in a case without the bow shock the equilibrium between
the ram pressures of the stellar wind and the ISW gives characteristic
dimensions of the interaction. For the ISW, velocity perpendicular
to the debris ring holds that the plane comprising the debris ring mid-plane
has the polar angle $\theta$ $=$ 90$^{\circ}$ in Eq. \ref{rBS}.
The shape of the bow shock varies in such away that the stellocentric
distance to the bow shock in the debris ring mid-plane is
$r_{\text{BS}}$ $=$ $r_{\text{BS}~0}$ $\sqrt{3}$.
In Fig. \ref{fig:BS} are shown $r_{\text{BS}}$ for number densities and
speeds corresponding in Fig. \ref{fig:density} to the dust particle with
$\bar{Q}'_{\text{pr~max}}$ $=$ 1.663. The stand-off distances determined
by the equilibrium between the ram pressures of the stellar
wind and the supersonic ISM gas are much smaller than that
of the debris ring. We must note that the theory in \citet{johnstone}
is obtained from a sample of main sequence stars. Since HD 61005 is
a pre-main sequence star, then the mass loss rate and the stellar wind
speed may not be accurately determined. However, as already mentioned,
the theory in \cite{johnstone} can be easily generalized also for
the pre-main sequence stars. The calculated values of $r_{\text{BS}}$
for the ISW parameters giving the observed shape morphology allow even
larger mass-loss rates and stellar-wind speeds to place the bow
shock inside the debris ring. Moreover, in the perpendicular case the bow
shock should cross the debris ring mid-plane with an inclination
$\arctan ( 4 / \pi )$ $\approx$ 51.9$^{\circ}$ given by the shape in
Eq. (\ref{rBS}). This means that if the debris ring is located in the bow shock,
the particles will leave the bow shock quickly on their trajectories giving
the observed morphology. In other words, the slope of the bow shock boundary
at the debris ring mid-plane is too high for the dust particle trajectories
to stay in the bow shock. The calculated stellocentric distances of the bow
shock in debris ring mid-plane ($r_{\text{BS}}$) are significantly smaller
than the stellocentric distances of the debris ring. Therefore, the debris
ring is more probably located above the bow shock. Since we assume that
the dust particles originate in the debris ring, the velocities and densities
of the ISW should not be affected by the interaction of the ISW with
the stellar wind.

\begin{figure}[t]
\begin{center}
\includegraphics[width=0.44005847953216374269005847953216\textwidth]{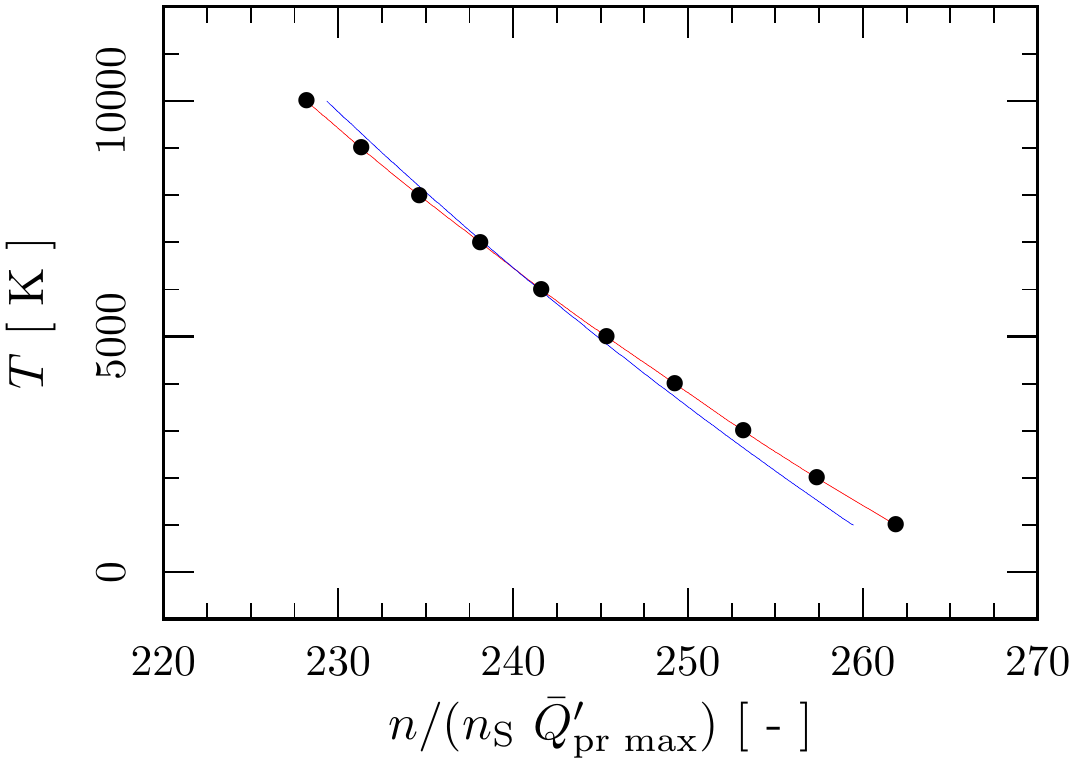}
\end{center}
\caption{Component density ratios giving the observed shape morphology at
HD 61005 for the ISW with speed 30 km/s and gas temperatures in the range
1000-10000 K (red line). The component density ratios are divided by maximal
averaged efficiency factor for the radiation pressure in the population
of dust particles. The blue line is used for the density ratios obtained
in the Stark approximation from the dependence of drag coefficients on
the temperature of the ISM gas (see Eq. \ref{cd}).}
\label{fig:temperature}
\end{figure}

The interaction of a stellar wind with an ISM gas creates
a termination shock in the stellar wind. The termination shock
in the Solar system was directly observed by {\it Voyager} 1
at a heliocentric distance 94 AU in 2004 and by {\it Voyager} 2
at 84 AU in 2007 \citep{richardson}. Stellocentric distance
to the termination shock is always smaller than that of the bow
shock for a given polar angle $\theta$ \citep[see e.g.,][]{frisch}.
Since the stellocentric distances to the bow shock in the debris ring
mid-plane at HD 61005 are smaller than 55-65 AU, the acceleration caused
by the stellar wind is ignored in our model.

\begin{figure*}
\begin{center}
\includegraphics[width=0.8\textwidth]{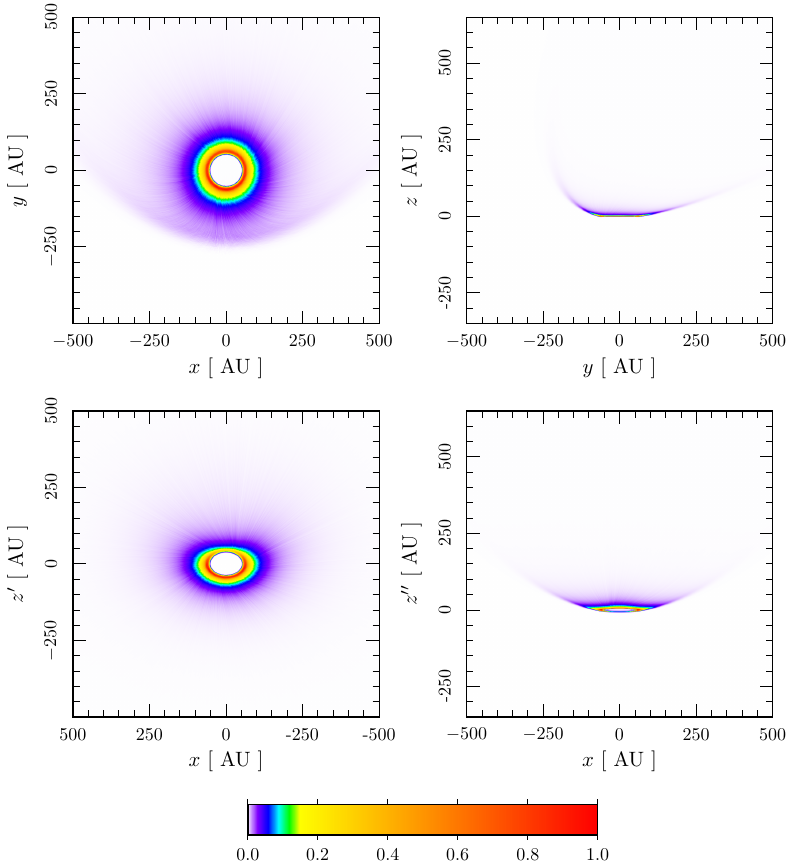}
\end{center}
\caption{The same situation as in Fig. 1 only with different interstellar
gas velocity vector and different viewpoints. The interstellar gas
velocity vector lies in the $xz$-plane and has 45$^{\circ}$ angle
between its direction and debris ring axis. Its magnitude
was modified using Eq. (\ref{tilted}) in order to obtain
the same observed morphology from the Earth. The bottom-left
plot shows the view on the star against the ISW.
The bottom-right plot shows the view from the Earth with the same
orientation as in Fig. \ref{fig:usable}.}
\label{fig:tilted}
\end{figure*}

\subsection{Temperatures of the ISW}
\label{subsec:temperature}

In Fig. \ref{fig:temperature} are shown temperatures and component
density ratios that give the observed morphology for the ISW speed 30 km/s
(red line). Large variation in gas temperature requires only small
variation in the density ratios in order to reproduce the observed morphology.
This is a consequence of properties of the drag coefficient. For the specular
reflection the dependence on the temperature is inside the Mach number and
obtained Mach numbers are sufficiently large to set the drag coefficients
close to 1. Therefore, the variations in the gas temperature do not
have a large influence on variation of the acceleration caused by the ISW.
A relation obtained from the Stark approximation is also shown
(blue line). Component density ratios for the Star approximation are given
by $\sum_{i = 1}^{3} c_{0i} n_{i} m_{i} / \bar{Q}'_{\text{pr~max}}$
$=$ constant $\approx$ 1.2853 $\times$ 10$^{-19}$ kg.m$^{-3}$, where
$c_{0i}$ $=$ $c_{Di}(s_{0i})$ and $s_{0i}$ $=$
$\sqrt{m_{i} / 2 k T_{i}} ~v_{\text{F}}$. The constant corresponds
to $T$ $=$ 6300 K for $v_{\text{F}}$ $=$ 30 km/s (the situation
depicted in Fig. \ref{fig:usable}). As in Fig. \ref{fig:density}
the simplified relation was obtained only with the acceleration from
the ISM gas without dependence on the particle's velocity.
An applicability of the Stark approximation was verified to be usable for
all ISW speeds measured in Fig. \ref{fig:density} in the temperature
range 1000-10 000 K.

\subsection{Diffuse reflection at the surface}
\label{subsec:diffusion}

For a diffuse reflection, the drag coefficient depends on the temperature
of the dust particle. At 60 AU from HD 61005, we obtain an equilibrium
temperature of the dust particle $\approx$ 31 K. The equilibrium
temperature should decrease with the stellocentric distance.
The maximal variation of the drag coefficient due to the diffusion is
according to Eq. (\ref{cd})
\begin{equation}\label{diffusion}
\left ( \frac{T_{\text{d}}}{T_{i}} \right )^{1 / 2}
\frac{\sqrt{\pi}}{3s_{i}} = \frac{\sqrt{\pi}}{3} \frac{1}{U}
\sqrt{\frac{2 k T_{\text{d}}}{m_{i}}} ~.
\end{equation}
For the dust particle at 60 AU the substitution of hydrogen with
the relative velocity 30 km/s in Eq. (\ref{diffusion}) yields the maximal
variation of the drag coefficient $\approx$ 0.014. Hence, the influence
of the diffusion reflection on the shape morphology should not be large.
This conclusion was verified also by the numerical solution of the equation
of motion with the diffuse reflection at the surface of the dust particles.

\subsection{ISW velocity tilted from debris ring axis}
\label{subsec:tilted}

The interstellar matter velocity vector may not be perpendicular to
the debris ring. Such a configuration can also produce what we observe
at HD 61005. Fig. \ref{fig:tilted} shows reproduced shape morphology
for the case where the angle between the ISM velocity vector
and the debris ring axis is 45$^{\circ}$. The ISM velocity vector
lies in the plane comprising the line of sight and the ring axis.
The bottom-left plot depicts the view against the ISW with the star
in the center. The bottom-right plot is created by the rotation
of the viewpoint in the top-left plot by 95.7$^{\circ}$
about the $x$-axis counterclockwise in the $yz$-plane in order to
show the shape morphology observed from the Earth similarly to
Fig. \ref{fig:usable}. We have used the same temperature and number
densities of the ISM gas components as in Fig. \ref{fig:usable}.
However, simple usage of the same speed of the ISM does not give
the observed morphology. To match observed morphology at this tilt
of the ISM velocity vector with respect to the debris ring axis the speed
must be increased. Using the same approximation as for the independence
of observed morphology on the particle's radius we can derive an equation
that gives how the speed of the ISM should be modified in order to obtain
the same observed morphology. A variation of the particle's position
caused by the Stark acceleration along the ISM velocity vector is
$z_{\text{S}}$ $=$ $C v_{\text{F}}^{2} t^{2} / 2$. Particle velocity
perpendicular to the ISM velocity vector is conserved in the Stark
approximation. Therefore, we can calculate time as $t$ $=$
$x_{\text{S}} / v_{\text{P}}$, where $x_{\text{S}}$ is the variation
of the particle's position in a direction perpendicular to
the ISM velocity vector and $v_{\text{P}}$ is the speed
of the particle in this direction. For the ISM velocity vector
lying in the plane comprising the line of sight and the ring axis we
obtain for two different configurations the same observed morphology if
$z_{\text{SA}} \sin \epsilon_{\text{A}}$ $=$
$z_{\text{SB}} \sin \epsilon_{\text{B}}$
and $x_{\text{SA}}$ $=$ $x_{\text{SA}}$.
Here, the subscripts A and B refer to two different configurations
and $\epsilon_{\text{A}}$ is an angle between the line of sight and
the ISM velocity in the configuration A. The modification equation
given by a substitution is
\begin{equation}\label{tilted}
\frac{v_{\text{FB}}^{2} \sum_{i = 1}^{N} c_{0i\text{B}} n_{i\text{B}}
m_{i\text{B}}}{v_{\text{F\text{A}}}^{2} \sum_{i = 1}^{N} c_{0i\text{A}}
n_{i\text{A}} m_{i\text{A}}} \approx
\frac{\sin \epsilon_{\text{A}}}{\sin \epsilon_{\text{B}}} ~.
\end{equation}

When the ISM velocity vector is tilted with respect to the ring axis,
the symmetry between the left and right ``wings'' observed from
the plane comprising the ISM velocity vector and the ring axis
is slightly affected. The slope of the ``wings'' slightly depends on
whether the motion of parent bodies in the debris ring is prograde
or rectograde. This is due to the fact that components of the initial
dust velocity parallel and perpendicular to the ISM velocity vector
vary along the mid-plane of the debris ring.

\subsection{Observed morphology of ISM particles in hyperbolic orbits}
\label{subsec:interdust}

 \citet{artcla} derived an equation of an axisymmetric
paraboloidal void caused by the stellar gravity reduced by the Keplerian
term from the PR effect in a flow of interstellar dust particles
approaching the star in hyperbolic orbits. For the dust particles with
constant $\beta$ the observed morphology should be similar to
tilted paraboloid $z$ $=$ $(x^{2} + y^{2}) / 4$ $-$ 1.
For a boundary of the paraboloid in an orthogonal projection to
a sky plane we obtain $z^{\prime}$ $=$ $( x^{2} / 4$ $-$
$1 / \sin^{2} \epsilon )$ $\sin \epsilon$. Here $\epsilon$ is the angle
between the line of sight and the ISM velocity ($\epsilon$ $\neq$ 0).
At a latitude of the star ($z$ $=$ 0) we obtain a slope of the boundary of
45$^{\circ}$, independently of $\epsilon$. This result can be easily applied
for a verification if an observed structure is caused by the dust
particles entering the star with the ISM. For an existing application
see \citet{gaspar}.

\section{Conclusion}
\label{sec:conclusion}

We have investigated the influence of ISW on the shape morphology
of an observed swept-back structure originating in the debris ring
at HD 61005. For explanation of the observed morphology, high
velocities or densities of interstellar gas are required.
On the boundary of the swept-back structure, the dust particles that are
maximally influenced by stellar radiation should by present.
When \citet{debes} was published, no observations of the debris
ring producing the dust particles existed. The assumed stellocentric
distances of about 10 AU used for the location of the debris ring were
much lower in comparison with today's real observed values of 55-65 AU.
Unfortunately, we were not able to reproduce \citet{debes} results using
the debris ring at 10 AU interacting with an interstellar hydrogen with
number density $n_{\text{H}}$ $=$ 100 cm$^{-3}$. It is likely that low assumed
stellocentric distance together with other reasons (see
Sect. \ref{subsec:density}) lead to higher initial velocities
of dust particles and consequently to higher densities and velocities
of ISW required for creation of the observed shape. Required
number density for the ISM with speed 25 km/s is 327.1 times higher
than the ISM entering the Solar system for the observed debris ring
in comparison with the value of 500 obtained for the interstellar hydrogen
in \citet{debes} for the debris ring at 10 AU. For such a strong
ISW the bow shock should be inside the debris ring.

The observed shape can be obtained also in the case where ISW
is not perpendicular to the ring plane. Suitable configuration
exists also in the case where the ISW velocity lies in a plane
comprising the line of sight and the debris ring axis. A simple
relation between suitable ISM speed and the angle between the line
of sight and ISM velocity can be obtained.

Gas temperature does not have a large influence on the shape
morphology due to larger values of Mach numbers that set the drag
coefficients close to 1.

In general, we can conclude that the observed morphology created
by the dust originating in the debris ring at HD 61005 can be
explained if we consider dust under the action of the ISW.

\begin{acknowledgements}
I am grateful to the Nitra Self-governing Region for support. I would
also like to thank the referees for suggestions which helped me improve
the manuscript.
\end{acknowledgements}

\begin{appendix}

\section{Initial elements after an ejection to hyperbolic orbit}
\label{appendix}

The positions and velocities of the parent body and the particle are
equal in a moment of the ejection with zero ejection velocities.
After an ejection from a parent body in an elliptical orbit
the semimajor axis of the particle orbit can be calculated
as follows. For the parent body we have
\begin{equation}\label{parent}
\frac{v^{2}}{2} - \frac{\mu}{r} = - \frac{\mu}{2 a_{\text{p}}} ~,
\end{equation}
where $v$ is the speed in the moment of ejection and $a_{\text{p}}$
is the semimajor axis of the parent body. The subscript $\text{p}$ will be
used for quantities belonging to the parent body. For dust particle with
$\beta$ $>$ 1 ejected to a hyperbolic orbit we have
\begin{equation}\label{particle}
\frac{v^{2}}{2} - \frac{\mu \left ( 1 - \beta \right )}{r} = -
\frac{\mu \left ( 1 - \beta \right )}{2 a} ~.
\end{equation}
The substitution of $v^{2}/2$ from Eq. (\ref{parent}) to Eq. (\ref{particle})
gives for the semimajor axis of the hyperbolic orbit
\begin{equation}\label{axis}
a = \frac{\left ( \beta - 1 \right ) a_{\text{p}}}
{2 \beta \cfrac{a_{\text{p}}}{r} - 1} ~.
\end{equation}

Comparison of the angular momentum $H$ written using the orbital
elements of the parent body and the particle in the moment of ejection
gives
\begin{align}\label{momentum}
H = {} & \left | \vec{r} \times \vec{v} \right | =
\sqrt{\mu a_{\text{p}} \left ( 1 - e_{\text{p}}^{2} \right )}
\notag \\
= {} & \sqrt{\mu a \left ( \beta - 1 \right ) \left ( e^{2} - 1 \right )} ~.
\end{align}
If we substitute Eq. (\ref{axis}) to Eq. (\ref{momentum}), then we can
determine the eccentricity of the hyperbolic orbit
\begin{equation}\label{eccentricity}
e^{2} = 1 + \frac{a_{\text{p}} \left ( 1 - e_{\text{p}}^{2} \right )}
{a \left ( \beta - 1 \right )} ~.
\end{equation}

The particle moves in the same plane as the parent body after
the ejection with the zero ejection velocity. Hence, the ascending nodes
and inclinations are equal
\begin{equation}\label{plane}
\Omega = \Omega_{\text{p}} ~, ~~ i = i_{\text{p}} ~.
\end{equation}

In the moment of ejection, the true longitudes of the parent body and
the particle are equal. Therefore,
\begin{equation}\label{longitudes}
f + \omega = f_{\text{p}} + \omega_{\text{p}} ~,
\end{equation}
where $f$ is the true anomaly and $\omega$ is the argument of pericenter.
For stellocentric distance of the dust particle with $\beta$ $>$ 1
on the hyperbolic orbit we have
\begin{equation}\label{radius}
r = \frac{a \left ( e^{2} - 1 \right )}{e \cos f - 1} ~.
\end{equation}
Radial components of velocities of the parent body ($v_{\text{Rp}}$) and
the particle ($v_{\text{R}}$) are equal in the moment of ejection
\begin{align}\label{radial}
v_{\text{Rp}} = {} &
\sqrt{\frac{\mu}{a_{\text{p}} \left ( 1 - e_{\text{p}}^{2} \right )}}
e_{\text{p}} \sin f_{\text{p}}
\notag \\
= {} & \sqrt{\frac{\mu \left ( \beta - 1 \right )}
{a \left ( e^{2} - 1 \right )}} e \sin f = v_{\text{R}} ~.
\end{align}
From the equality of transversal components of the velocities we obtain
\begin{align}\label{transversal}
v_{\text{Tp}} = {} &
\sqrt{\frac{\mu}{a_{\text{p}} \left ( 1 - e_{\text{p}}^{2} \right )}}
\left ( 1 + e_{\text{p}} \cos f_{\text{p}} \right )
\notag \\
= {} & \sqrt{\frac{\mu \left ( \beta - 1 \right )}
{a \left ( e^{2} - 1 \right )}} \left ( e \cos f - 1 \right ) =
v_{\text{T}} ~.
\end{align}
Using Eq. (\ref{momentum}) in Eqs. (\ref{radial}) and (\ref{transversal})
we can calculate the true anomaly
\begin{equation}\label{atan2}
e \sin f = \frac{e_{\text{p}} \sin f_{\text{p}}}{\beta - 1} ~,~~
e \cos f = \frac{\beta + e_{\text{p}} \cos f_{\text{p}}}{\beta - 1 } ~.
\end{equation}
The true anomaly substituted in Eq. (\ref{longitudes}) determines the argument
of pericenter. For particles with $\beta$ $\approx$ 3 released from parent
bodies with nearly circular orbits ($e_{\text{p}}$ $\approx$ 0) the true
anomaly is always close to zero, in other words, the particle is ejected close
to the pericenter of the hyperbolic orbit.

\end{appendix}

\end{document}